\documentclass[cits]{PoS}

\title{New results on the parsec-scale properties of the Bologna Complete Sample}

\author{\speaker{Gabriele Giovannini}$^{a,b}$, Elisabetta Liuzzo$^{a,b}$,
        Marcello Giroletti$^b$, and Greg B. Taylor$^c$\\
        \llap{$^a$} Dipartimento di Astronomia di Bologna\\
                   via Ranzani 1, 40126 Bologna (I)\\
        \llap{$^b$} Istituto di Radioastronomia - INAF\\
                   via Gobetti 101, 40129 Bologna (I)\\
        \llap{$^c$} University of New Mexico\\
                    Albuquerque - NM (USA)\\
        E-mail: \email{ggiovann@ira.inaf.it}, \email{liuzzo@ira.inaf.it},
                 \email{m.giroletti@ira.inaf.it}, \email{gbtaylor@unm.edu}}

\abstract{We present an update of the parsec scale properties of the Bologna 
Complete Sample consisting of 95 radio sources with z $<$ 0.1 from the B2 
Catalog of Radio Sources and the Third Cambridge Revised Catalog (3CR).
Thanks to new data obtained in phase reference mode, we have now parsec scale 
images for 76 sources of the sample. Most of them show a one-sided jet 
structure but we find a higher fraction
of two-sided sources in comparison with previous flux-limited VLBI
surveys. Results for two peculiar
sources, 3C 293 and 3C 310, are presented and discussed in more detail.
}

\FullConference{The 9th European VLBI Network Symposium on The role of VLBI in the Golden Age for Radio Astronomy and EVN Users Meeting\\
		 September 23-26, 2008\\
		 Bologna, Italy}

\ShortTitle{The Bologna Complete Sample}

\begin{document}

\section{Introduction}

The study of the parsec scale properties of radio galaxies is crucial to
obtain information on the nature of their central engine.
In order to study statistical properties of different classes of sources, it
is necessary to define and observe a sample that is free from selection
effects. To this aim, it is important to define the sample using a low radio
frequency, to avoid observational biases related to orientation effects,
thanks to the dominance at low radio frequency of the unbeamed extended
emission. With this purpose, we started a project to observe a complete
sample of radio galaxies (the Bologna Complete Sample (BCS))
selected within the B2 Catalog of Radio Sources and
the Third Cambridge Revised Catalog (3CR) \cite{gio90, gio01},
with no selection constrain on the nuclear
properties.

The sample consists of 95 radio sources. We selected
the sources above a homogeneous flux density limit of 0.25 Jy at 408 MHz for
the B2 sources and those above 10 Jy at 178 MHz for the 3CR sources
\cite{fer84} and applied the following criteria: 1) declination 
$>$10$^{\circ}$; 2) Galactic latitude $|$ b $|$ $>$ 15 $^{\circ}$;
redshift z$<$0.1.

At present VLBI observations are available for 76 sources. The missing 19
sources will require very sensitive observations (phase reference technique
and large bandwidth) because of the very low activity of the radio
core.

\section{Source morphology}

At parsec scale most of the sources (24 FR I and 7 FR II) show as expected a
one-sided jet structure because of relativistic effects, however we have
also a high number of sources with a symmetric jet structure.
We classify as two
sided all sources showing both a jet and counterjet, regardless of the value
of the jet to the counterjet ratio or the length of the counterjet.
The total number of sources with a two-sided morphology is 17, corresponding
to $\sim$ 25$\%$. This percentage
 is significantly higher than that found in previous samples in the
literature:
there are 11$\%$ symmetric sources in the PR (Pearson-Readhead) sample
\cite{pea88} and 4.6$\%$ (19/411) in the
combined PR and
Caltech-Jodrell (CJ) samples \cite{tay94,pol95,pec0}.

For two-sided sources, the brightness ratio between the two jets is in
the range 1--5, while most of the brightness lower limit in one-sided
sources is higher than 5 confirming that the
sensitivity in our maps is generally good enough to detect two sided sources.
The main difference between the percentage of symmetric sources in the
present sample and in previous samples is naturally explained in the framework
 of unified scheme models since our sources have been
selected at low frequency and should not be
affected by an orientation bias.
The comparison between the VLA and VLBA jet position angles (P.A.)
shows that most of the
sources do not
show a large misalignment and that
the one-sided parsec scale (or the brigther jet in double-sided parsec jets)
is oriented with respect to the nuclear emission, on the same side of the
brighter kiloparsec-scale.

We compared the total flux at VLBA scale with the core arcsecond flux density.
We find that 70 $\%$ have a
correlated flux density larger than 80$\%$ of the arcsecond core flux density.
 This means that in these sources we mapped most of the milliarcsecond (mas)
scale structure
and so we can properly connect the parsec to the kiloparsec structures.
For 19 (30$\%$) sources, a significant fraction of the arcsecond core
flux density is missing in the VLBA images. This suggest variability or  the
presence of significant structures between $\sim$30 mas and 2 arcsec that the
VLBA can miss because the lack of short baselines.
To properly study these structures, future observations
with EVLA or the e-MERLIN array will be necessary.

\section{Individual sources}

{\bf 1350+31 -- 3C293}
This peculiar source has been studied in detail in the radio and optical
bands. Recent results are discussed in \cite{bes04} and \cite{flo06}, 
where the source structure is analyzed from the sub-arcsec to the
arcmin scale. In Giovannini et al. 2005 (\cite{gio05}, 
we presented a VLBI image at 5 GHz
where a possible symmetric two-sided jet structure is visible at mas
resolution. Because of the complex inner structure we observed
again this source with the VLBA at 1.7 GHz in phase reference mode to properly
map the inner region. Moreover we reduced VLA data in the A+ configuration
(A configuration, adding also 
the VLBA-Pt telescope) to increase the VLA angular resolution and to study the
connection between the sub-arcsec and the arcsec structure.
In Fig. 1  we show the VLBI image of the core. 

\begin{figure}
\includegraphics[width=.5\textwidth]{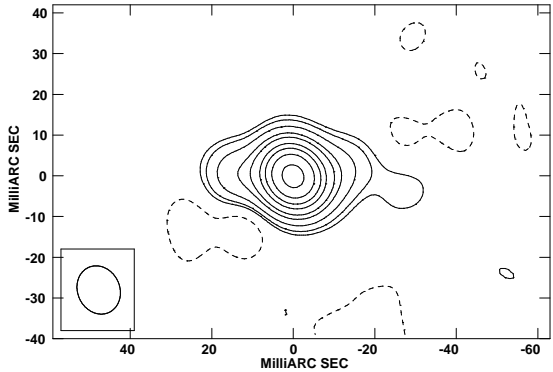}
\caption{VLBA image at 1.7 GHz of 3C 293. Contours are: -0.3 0.3 0.5
1 2 3 5 7 10 15 mJy/beam.}
\label{fig1}
\end{figure}

In agreement with the previous
VLBA image at 5 GHz, the source shows a symmetric structure with two 
symmetric jets.
The sub-arcsec VLA image
at 5 GHz shows the bright inner two-sided jet with a more external fainter 
diffuse
emission (Fig. 2). 
\begin{figure}
\includegraphics[width=.5\textwidth]{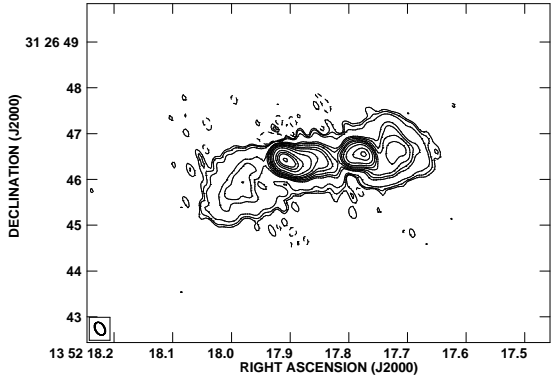}
\caption{VLA+ (VLA in A configuration and VLBA-Pt) image at 5  GHz of 3C 293.
Levels are: -0.3 0.3 0.5 1 5 7 10 15 20 30 50 100 150 200 300 mJy/beam.}
\label{fig2}
\end{figure}
The brightness distribution suggests 
a two
phase activity: the source emission after some time re-started with
a higher activity with respect to the older emission.
However the jet direction appears to
be constant in between the two activity epochs: after some oscillations
the subarcsec jet is clearly oriented in E-W direction, and it moves to 
a PA = -20 (E side)
and to a PA + 20 (W side) in the direction of the arcsecond scale jet 
(Fig. 3).

\begin{figure}
\includegraphics[width=.6\textwidth]{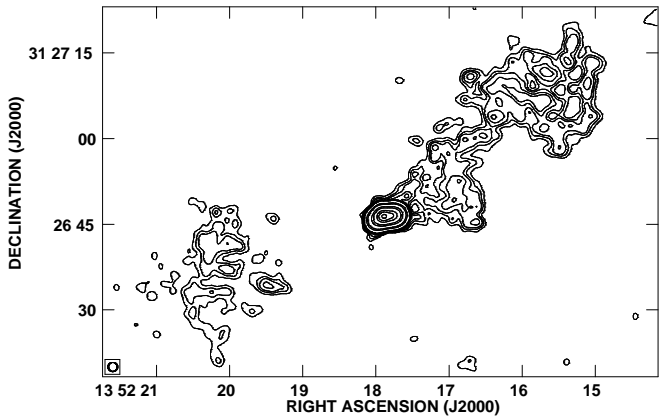}
\caption{VLA image at 5 GHz (B and C configuration) of 3C 293. Levels are:
0.1 0.2 0.3 0.5 0.7 1 5 10 50 100 300 500 800 mJy/beam)}
\label{fig3}
\end{figure}

Our suggestion is that this source is at a large angle with respect to the
line of sight, and that oscillations
and relatively small changes in the jet PA are present. In the region at
about 2 - 3 kpc from the core the Jet PA shows a large symmetric change
with a kpc scale structure almost in NE-SE direction. 
This large change
is not due to a change in the restarting nuclear activity but it looks
constant in time and it is likely produced by the jet interaction
with a rotating disk as discussed in \cite{breu84}.

{\bf 1502+26---3C310} 
It is identified with a 15 mag elliptical galaxy at
z = 0.0538. In optical images
this galaxy is flattened east-west, on an almost perpendicular direction to 
the radio jet axis. 
In the radio band at low resolution it appears as
a relaxed double (\cite{breu84b}) with a steep radio spectrum
($\alpha^{26}_{750}$ = 0.9; $\alpha^{0.75}_{10.7}$ = 1.4). 
\begin{figure}
\includegraphics[width=.6\textwidth]{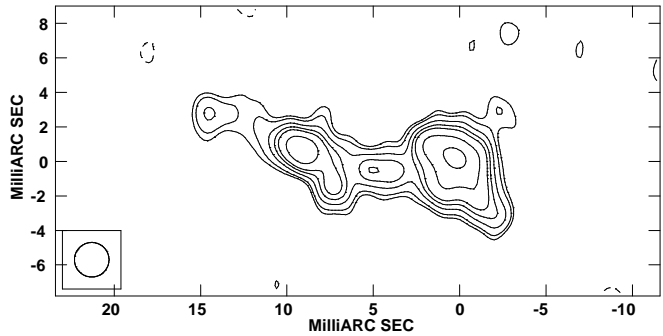}
\caption{VLBA image at 5 GHz of 3C 310. Levels are: -0.2 0.15 0.2 0.3 0.4 0.7 
1 mJy/beam}
\label{fig4}
\end{figure}

\begin{figure}
\includegraphics[width=.5\textwidth]{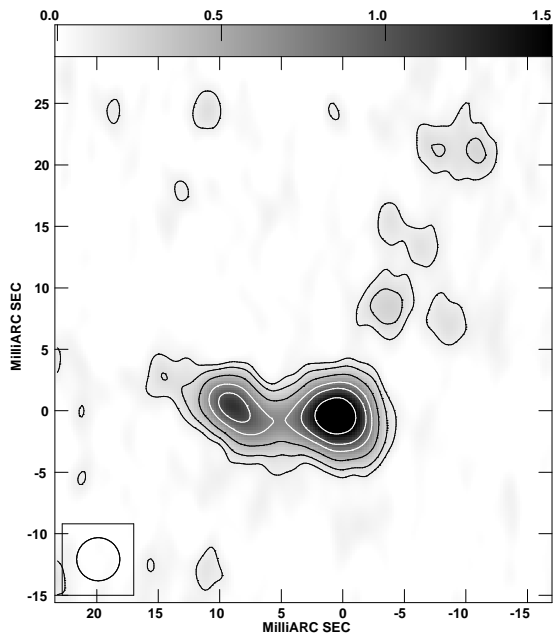}
\caption{Low resolution VLBA image at 5 GHz of 3C 310. Levels are:
0.1 0.2 0.4 0.7 1 1.5 mJy/beam.}
\label{fig5}
\end{figure}

\begin{figure}
\includegraphics[width=.4\textwidth]{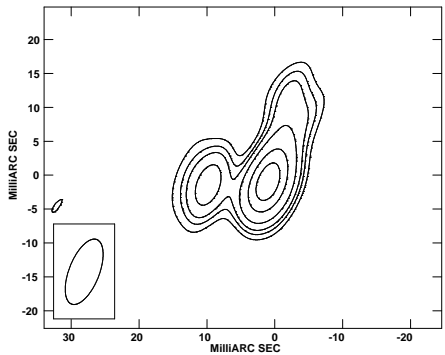}
\caption{VLBA image at 1.6 GHz of 3C 310. Levels are: -1 1 1.5 2 3 5 7 
mJy/beam.}
\label{fig4}
\end{figure}
Low
brightness lobes show a complex structure with filaments and bubbles.
There is an
unresolved radio core with a faint extension from the core to the North,
extended 5'', possibly identified as a faint jet (\cite{breu84b}).
The VLBA structure is very complex. In \cite{gio05} we showed an image
interpreted as a one-sided emission in contrast with \cite{giz02}.
Because of the short observation and low flux density we observed
again this source at 5 and 1.4 GHz in phase reference mode.
At 5 GHz we have an elongated structure in E-W direction with a possible 
central nuclear structure and two symmetric jets or mini-lobes (Fig. 4). 
At lower
resolution (Fig. 5) a faint emission is visible in the N direction, at the
sampe PA of the kpc scale structure.
At 1.4 GHz the E-W
structure is confirmed (Fig. 6), as well as the extension to the North 
in agreement with \cite{giz02}.
We produced images at 5 and 1.4 GHz at the same resolution and cellsize and
obtained a spectral index image. The spectral index distribution
confirms the identification of the core with the faint emission at the center
of the pc scale structure which shows a symmetric two-sided extension with a
very steep spectral index. 

The low resolution images suggest that after a few
mas there is large change in the source PA 
from E-W to N-S in agreement with the kpc scale. This peculiar morpholoy 
could be related to the interaction with a rotating disk as in 3C293 or to
the presence of a double nucleus. In any case the pc scale structure 
visible in Fig. 4 and the large change in the radio structure PA suggest
a sub-relativistic velocity for the two-sided emission.

\acknowledgments 
The National Radio Astronomy
Observatory is operated by Associated Universities, Inc., under cooperative
agreement with the National Science Foundation.

\end{document}